\documentclass[a4paper, 10pt, conference]{ieeeconf}      

\IEEEoverridecommandlockouts                              
\usepackage{graphics} 
\usepackage{epsfig} 
\usepackage{times} 
\usepackage{amsmath} 
\usepackage{amssymb}  

\title{\LARGE \bf
Compensation of Nonlinear Torsion in Flexible Joint Robots:
Comparison of Two Approaches}

\author{Michael Ruderman
\thanks{M. Ruderman is with Department of Electrical, Electronics, and Information Engineering,
Nagaoka University of Technology, 1603-1, Kamitomioka, Nagaoka,
940-2188 Japan {\tt\small ruderman@vos.nagaokaut.ac.jp}}%
}

\begin{document}

\maketitle
\thispagestyle{empty}
\pagestyle{empty}

\begin{abstract}
Flexible joint robots, in particularly those which are equipped
with harmonic-drive gears, can feature elasticities with
hysteresis. Under heavy loads and large joint torques the
hysteresis lost motion can lead to significant errors of tracking
and positioning of the robotic links. In this paper, two
approaches for compensating the nonlinear joint torsion with
hysteresis are described and compared with each other. Both
methods assume the measured signals available only on the motor
side of joint transmissions. The first approach assumes a
rigid-link manipulator model and transforms the desired link
trajectory into that of the motor drives by using the inverse
dynamics and inverse hysteresis map. The second approach relies on
the modeling of motor drives and inverse hysteresis and uses the
generalized momenta when predicting the joint torsion. Both
methods are discussed in details along with a numerical example of
two-link planar manipulator under gravity.
\end{abstract}


\bstctlcite{references:BSTcontrol}

\section{INTRODUCTION}
\label{sec:1}

Joint elasticities in robotic manipulators, see e.g.
\cite{Readman1994,spong2006,Sicil2009} for details, may provoke
the disturbing vibrations of the links but also the relative
torsion between the motor drives under control and joint output
axes. When the control in joint space operates using the motor
drive feedback only, that is the most common case in robotic
practice, the relative joint torsion remains uncompensated and
leads to the link position errors at heavy loads and large joint
torques. The measures of compensating the gravity-induced torsion
in robotic joints with linear elasticities have been elaborated
and reported in former works \cite{Tomei91}, \cite{spong87}.
However, when accounting for torsion-torque hysteresis, which is
the matter of fact in various geared manipulators and particularly
those equipped with harmonic drives, single gravity-related
compensation, like one in \cite{DeLucaEtAl05}, can be
insufficient. This becomes particularly visible when a high
positioning accuracy of the links is required. The experimental
evidence of torsion-torque hysteresis in a geared single joint
with harmonic-drive can be found in e.g.
\cite{Ruderman2014b,Ruderman2014c}. Further explicit studies of
nonlinearities in harmonic-drives can be found in
\cite{tuttle1996,taghirad98,Ghorbel2001,dhaouadi03}, and in the
context of robotic joints in the former works
\cite{Kircanski94,Seyfferth95}. From the last developments in
controller design suitable for elastic joint robots, the immersion
and invariance (I\&I) method \cite{Astolfi2007} can be further
mentioned. The method assumes, however, the state feedback
available, i.e. position and velocity also of the robot links
behind the gear transmission.

This paper makes use of the recently proposed and elaborated joint
torsion compensation based on the so-called `virtual sensor'
\cite{Ruderman2014b,Ruderman2014c,Ruderman2015,Ruderman2015Last}.
Two different approaches are described in the following and
compared with each other. The first one is inspired by the
feed-forward control law provided in \cite{DeLuca00}. In the
recent paper we extend the reference trajectory transformation and
feed-forward control law to the case of nonlinear joint torque
with hysteresis. The second approach, based on the previous works
\cite{Ruderman2014b,Ruderman2014c,Ruderman2015}, observes the
actual joint torque from the given motor drive signals and makes a
prediction of relative joint torsion by using the inverse
hysteresis map. The predicted relative joint torsion, and that
with low-pass characteristics, is augmented to the feedback motor
drive position, thus providing a 'virtual' sensing of link's
position. Both approaches are shown as being integrated into the
two-degrees-of-freedom control, including the model-based
feed-forwarding and proportion-derivative feedback. We show and
compare the control performance of both approaches by using the
numerical example of a standard two-link planar manipulator with
gravity.

\section{DYNAMICS OF ELASTIC JOINT ROBOTS WITH NONLINEARITIES}
\label{sec:2}

\begin{figure}[!h]
\centering
\includegraphics[width=0.45\columnwidth]{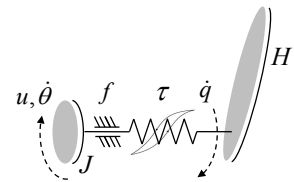}
\caption{Elastic robotic joint with hysteresis and friction
nonlinearities} \label{fig:joint}
\end{figure}

We consider the flexible joint robotic manipulator as
\begin{eqnarray}\label{eq:j1}
  H(q)\ddot{q} +C(q, \dot{q}) +G(q)  &=& \tau(\Delta,t), \\
  J\ddot{\theta} + \tau(\Delta,t) &=& u - f(\dot{\theta}),
\label{eq:j2}
\end{eqnarray}
where the left-hand side of the first equation describes the
standard rigid-link dynamics, and the second equation describes
the joint drives actuated by the vector of motor torques $u$. The
vector of angular coordinates of the link axes is denoted by $q$,
and the vector of angular rotor displacements of the motor drives,
reflected through the gear ratios, is denoted by $\theta$. The
relative angular displacement between both is denoted by $\Delta =
\theta - q$ and constitutes the vector of joint torsion. The
schematic representation of an elastic robotic joint described by
(\ref{eq:j1}), (\ref{eq:j2}) is illustrated in Fig.
\ref{fig:joint}.
Note that if assuming the linear joint stiffness, i.e. $\tau = K
\Delta$ where $K$ is a positive diagonal matrix, and neglecting
the nonlinear motor drive friction $f$, the model (\ref{eq:j1}),
(\ref{eq:j2}) reduces to the well-established model of elastic
joint robots initially proposed by Spong in \cite{spong87}. The
rigid-link dynamics is parameterized by $H(q) \in \mathbb{R}^{n
\times n}$ inertia matrix of manipulator and $C(q,\dot{q}) \in
\mathbb{R}^{n}$ and $G(q) \in \mathbb{R}^{n}$ vectors of
Coriolis/centrifugal and gravity torques correspondingly. $J =
\mathrm{diag} (j_{i}) \in \mathbb{R}^{n \times n}$ is the positive
diagonal matrix of motor drive inertias. The objective of our
model extension is to capture the nonlinear joint elasticities
with hysteresis, as will be shown further in Section III.
Heretofore we will keep, however, the joint torque $\tau$ as a
generic nonlinear function of relative torsion and time, hence
without loss of generality.

To keep the modeling of motor drives friction simple as possible
while capturing, at the same time, the most pronounced friction
nonlinearities we apply the steady-state Stribeck friction curve.
Note that a more complex (dynamic) friction behavior includes the
phenomenon of friction lag, also known as hysteresis in the
velocity, and the presliding hysteresis in displacement, see e.g.
\cite{ArmstDupontCan94}, \cite{AlBendSwev08} for details. The
total friction torque of the $i$-th motor
drive\footnotemark{\footnotetext{Note that here and further in (4)
we skip the index $i$ for the sake of simplicity, while a scalar
value, computation is meant.}} is described by
\begin{equation}\label{eq:f}
f\bigl(\dot{\theta}\bigr) = \mathrm{sig} (\dot{\theta})\Bigl(F_{c}
+ F_{s} \, \exp\bigl[ - V^{-\mu} \bigl|\dot{\theta}\bigr|^{\mu}
\bigr] \Bigr) + B \dot{\theta}.
\end{equation}
Here the standard Stribeck characteristic curve is parameterized
by the Coulomb and Stribeck friction coefficients $F_{c} > 0$ and
$F_{s} > 0$ correspondingly. The exponential parameters $\mu \neq
0$ and $V > 0$ are respectively the Stribeck velocity and shape
factors; both determine the velocity weakening and strengthening
curve. The linear viscous friction coefficient is denoted by $B$.
Note that the applied sigmoid function
\begin{equation}\label{g}
\mathrm{sig} (\dot{\theta}) = \frac{2}{1 + \exp(-\gamma
\dot{\theta})}-1
\end{equation}
allows avoiding the discontinuity at zero velocity crossing, while
$\gamma$ is the velocity scaling factor.

\section{REFERENCE TRAJECTORY TRANSFORMATION}
\label{sec:3}

The first approach relies on the model-based transformation
between the given link reference $q_{r}$ and motor-drive reference
$\theta _{r}$ which is provided to the feedback control.
Furthermore, the feed-forward control in the transformed $\theta
_{r}$ coordinates is used. As has been shown in \cite{DeLuca00},
the reference trajectory of the desired robot link position can be
transformed into that of the motor drives by using the inverse
model of rigid-link dynamics and inverse matrix of joint stiffness
coefficients. Since the joint elasticities are not longer linear
we are to extend the reference trajectory transformation to the
general case of a nonlinear torsion-torque map $\tau =
\chi(\Delta)$. Given the reference link trajectory $q_{r}(t) \in
\mathcal{C}^{4}$ one obtains the desired motor drive trajectory as
\begin{equation}\label{1}
\theta_{r} = q_{r} + \chi^{-1}\bigl(\tau_{r}\bigr),
\end{equation}
where the reference joint torque $\tau_{r}$ is computed according
to (1). Differentiating twice with respect to the time, whilst
taking into account the rate-independency of hysteresis, i.e.
$\frac{d}{dt} \partial \chi ^{-1}/\partial \tau = 0$, one obtains
\begin{equation}\label{2}
\ddot{\theta} _{r} = \ddot{q} _{r} + \frac{\partial \chi
^{-1}}{\partial \tau_{r}}  \ddot{\tau} _{r}.
\end{equation}
One can see that in order to realize the motor drive reference (6)
the corresponding joint link reference should be at least 4 times
differentiable since
\begin{equation}\label{3}
\ddot{\tau} _{r} = H(q_{r}) q_{r}^{(4)} +
2\dot{H}(q_{r})q_{r}^{(3)} + \ddot{H}(q_{r})\ddot{q}_{r} +
\ddot{C}(q_{r}, \dot{q}_{r}) + \ddot{G}(q_{r}).
\end{equation}
Now, having the reference trajectory transformation, and with
respect to the motor drive dynamics (2), one obtains the reference
feed-forward control as
\begin{equation}\label{4}
u_{r} = J\ddot{\theta} _{r} + \chi(\theta_{r} - q_{r}) +
f(\dot{q}_{r}) \equiv J\ddot{\theta} _{r} + \tau_{r} +
f(\dot{q}_{r}).
\end{equation}
Note that the friction compensating term is included in (8) while
assuming $f(\dot{q}) \approx f(\dot{\theta})$ for the reference
value. This is justified at least for non-zero velocities, i.e.
apart from the motion reversals.

\section{OBSERVATION OF JOINT TORSION}
\label{sec:4}

The second approach relies on the fact that the reactive joint
torque appears as an input disturbance of the motor drives under
control. Assuming the model of motor drives is given and the
input-output data tuples $(u+\tau, \dot{q})$ are available from
the measurement, the input disturbance $\tau$ can be detected and
isolated by using the method of so-called generalized momenta, see
\cite{DePersis01,DeLuca03} for details.

Introducing the vector of generalized momenta $p = J \dot{\theta}$
we rewrite the motor drive dynamics (2) as
\begin{equation}\label{eq:6}
\dot{p} = u - f(\dot{\theta}) - \tau.
\end{equation}
Since the drive and friction torques are known from the
measurement, the dynamics of generalized momenta can be estimated
by
\begin{equation}\label{eq:7}
\dot{\tilde{p}} = u - f(\dot{\theta}) - r,
\end{equation}
where the residual vector $r = L(\tilde{p} - p)$ is proportional
to the estimation error. The diagonal matrix $L > 0$ is a design
parameter. Note that (\ref{eq:7}) is a standard observer for the
class of dynamic systems with nonlinear term, here friction, as a
function of measurable outputs \cite{DeLuca03}. Further it can be
shown that since
\begin{equation}\label{eq:8}
r = L \Bigl( \int \bigl[ u - f(\dot{\theta}) - r \bigr]dt - p
\Bigr),
\end{equation}
the residual state dynamics complies with
\begin{equation}\label{eq:9}
\dot{r} + Lr = L \tau.
\end{equation}
It is evident that the residual state $r$ follows the unknown
joint torque and, by doing this, exhibits the first-order time
delay behavior with the time constants $L^{-1}$.

The detected and isolated reactive joint torque $\tilde{\tau} = r$
serves as the input of inverse hysteresis model
\begin{equation}\label{5}
\tilde{\Delta} (\tilde{\tau}) = \alpha ^{-1} \Bigl( W^{-1}
\bigl[\tilde{\tau} - (I - W)\beta (\tilde{\Delta})\bigr]\Bigr),
\end{equation}
where $\alpha$ and $\beta$ are the static and dynamic terms of the
Bouc-Wen-like hysteresis model, see \cite{Rud12a,Ruderman2014b}
for details. $W = \mathrm{diag}(w_i) \in \mathbb{R}^{n \times n}$
is the diagonal matrix of weighting factors $0 < w _{i} < 1$ and
$I$ is the identity matrix. The static term of Bouc-Wen-like
hysteresis model is given by
\begin{equation}\label{6}
\bar{\tau}_i = \alpha (\Delta_i) = k_{1} \Delta_i + k_{3}
\Delta_i^{3},
\end{equation}
and captures the stiffening spring characteristics. The dynamic
term
\begin{equation}\label{7}
\hat{\tau}_i = \beta(\Delta_i) = k_{1} \int \dot{x}_i dt + k_{3}
\Bigl(\int \dot{x}_i dt\Bigr)^{3}
\end{equation}
with an internal state
\begin{equation}\label{8}
\dot{x}_i = \dot{\Delta}_i - \psi |\dot{\Delta}_i| |x_i|^{\eta-1}
x_i -\xi \dot{\Delta}_i |x_i|^{\eta}
\end{equation}
captures the actual hysteresis state and is parameterized by the
hysteresis control parameters $\psi$, $\xi$, and $\eta$. The
parameters $k_{1}$ and $k_{3}$ are the linear and cubic stiffness
coefficients. Note that the relationship between a purely elastic
and purely plastic (hysteresis) contributions is determined by the
weighting factor $w$. Assuming the Bouc-Wen-like torsion-torque
hysteresis is given (identified) and reactive joint torque is
observed, the vector of joint torsion can be computed online by
(13). For more details on this method, also denoted as virtual
sensor of joint torsion, and its experimental evaluation the
reader is referred to \cite{Ruderman2014b,Ruderman2014c}.

An independent joint control with the motor drive feedback and
virtual sensor of joint torsion is  given by
\begin{equation}\label{9}
u = K_{p}e + K_{d}\dot{e} + K_{p} \tilde{\Delta}(\tilde{\tau}),
\end{equation}
where $K_{p}$ and $K_{d}$ are the positive diagonal matrices of
proportion and derivative control gains, and $e = q_{r} - \theta$
is the control error in the joint link space. It is easy to
recognize that the proportional control part operates on the
$q_{r} - \theta + \tilde{\Delta}$ error quantity, and once
$\tilde{\Delta} = \Delta$ correspondingly on $q_{r} - q$.
Therefore the steady-state error of feedback control (17), in the
joint link space, will be the same as the residual error of
predicting the relative joint torsion. Important to note is that
since the $\tilde{\tau} / \tau$ transfer characteristics
constitutes a low-pass filter, the feedback of $\tilde{\tau}$,
mapped through the hysteresis function, cannot destabilize the
closed-loop system. Recall that the hysteresis map $\tilde{\tau}
\mapsto \tilde{\Delta}$ by itself serves as an additional
rate-independent damping, see e.g. \cite{Lazan68}. In the rest of
this Section, we will prove the stability assumption made above
for the weaker condition $\tilde{\Delta} = K^{-1} \tilde{\tau}$,
i.e. without additional hysteresis damping. In the following we
assume that the joint couplings and configuration-dependent
nonlinearities of manipulator dynamics are compensated by the
model-based feed-forwarding. Furthermore, we approximate the motor
drive friction by the linear viscous friction $f \approx B \,
\dot{\theta }$ which is appropriate except for the motion
reversals.

The linearized model of the $i$-th elastic robotic
joint\footnotemark{\footnotetext{Note that here and for the rest
of this Section we skip the index $i$ for the sake of simplicity,
while scalar value computations are meant.}} can be written in
Laplace $s$-domain as
\begin{eqnarray} \label{eq:19a}
  P_{l} (s) q(s)       &=& K \theta(s),\\
  P_{m} (s) \theta(s)  &=& u(s) + K q(s),
  \label{eq:19b}
\end{eqnarray}
where
\begin{eqnarray}
\label{eq:19c}
  P_{l} (s) &=& \hat{H} s^{2} + K,\\
  P_{m} (s) &=& Js^{2} + B s + K
  \label{eq:19d}
\end{eqnarray}
are the forward transfer functions of the joint links and motor
drives correspondingly. The forward and feedback couplings between
both are provided by the joint stiffness $K$. The linearized
(decoupled) inertia of manipulator links is denoted by $\hat{H}$.
When using the predicted joint torsion in feedback, i.e. $u =
K_{p} \tilde{\Delta}$, the motor drive equation (19) becomes
\begin{equation}\label{19d}
\Bigl(P_{m} - \frac{K_{p}}{L^{-1}s + 1}\Bigr) \theta(s) = u(s) +
\Bigl(K - \frac{K_{p}}{L^{-1}s + 1}\Bigr)q(s).
\end{equation}
Considering the open-loop system (18), (22) one can rewrite the
characteristic polynomial of the transfer function $ \theta(s)/
u(s)$ into the form
\begin{equation}\label{19e}
1 + K_{p} G(s) = 0,
\end{equation}
where
\begin{equation}\label{19f}
G(s) = \frac{K - (\hat{H}s^{2} + K)\bigl((L^{-1}s + 1)(Js^{2} + Bs
+K)-1\bigr)}{K^{2}(L^{-1}s +1)}.
\end{equation}
The root locus of characteristic polynomial (24) is exemplarily
shown in Fig. 2 together with the pole-zero diagram of the nominal
plant (18), (19). One can see that when increasing the feedback
gain $K_{p}$ the critical, i.e. from stability point of view, pole
moves towards the conjugate-complex zeros of nominal plant. Here,
a relatively large gain variation is possible without entering the
marginally-stable region close to imaginary axis. One should note
that the variation of $L$, which is equally a design parameter,
reshapes the root locus trajectories so that an admissible $K_{p}$
range will equally change. However, it is obvious that the
feedback of predicted joint torsion is not destabilizing the
transfer characteristics of elastic robotic joint. A suitable
trade-off between $L$ and $K_{p}$ can be found during the first
stage of the control design, i.e. before closing the feedback loop
by the PD term.
\begin{figure}[!h]
\centering
\includegraphics[width=0.95\columnwidth]{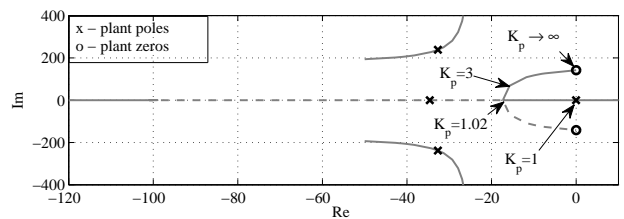}
\caption{Root locus (with respect to $K_p$) of elastic joint
extended by VS, i.e. (18), (22), versus pole-zero diagram of the
nominal plant (18), (19)} \label{fig:pzmap}
\end{figure}

\section{NUMERICAL EXAMPLE}
\label{sec:5}

\begin{figure}[!h]
\centering
\includegraphics[width=0.8\columnwidth]{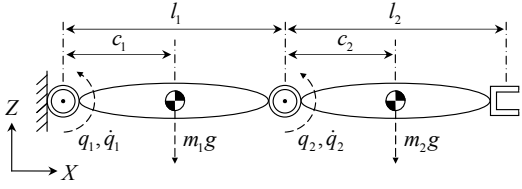}
\caption{Two-link planar manipulator under gravity}
\label{fig:planararm}
\end{figure}

\begin{figure}[!h]
\centering
\includegraphics[width=0.48\columnwidth]{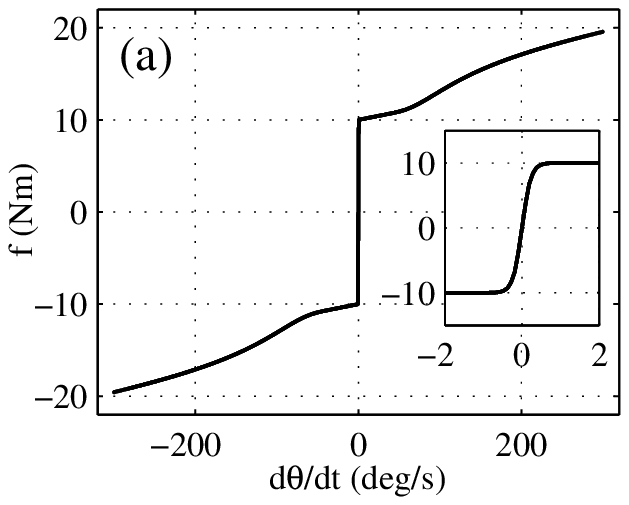}
\includegraphics[width=0.48\columnwidth]{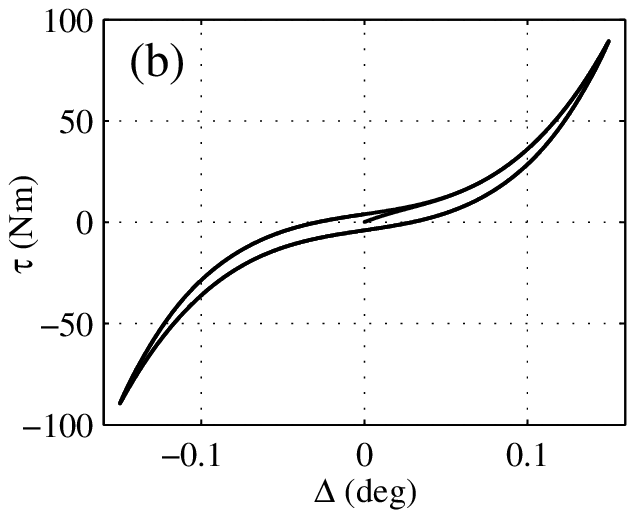}
\caption{Characteristic curves of nonlinear friction (a) and
hysteresis (b)} \label{fig:nonlincurves}
\end{figure}

Consider a classical example of two-link planar manipulator with
revolute joints under impact of gravity as shown in Fig.
\ref{fig:planararm}. Note that this structure is particularly
interesting since coinciding with the `shoulder' and `elbow' axes
of several anthropomorphic, also known as RRR, industrial robotic
manipulators.
The detailed deviation of kinematics and dynamics of a two-link
planar manipulator under gravity can be found in e.g.
\cite{spong2006}, \cite{Sicil2009}. Here we note that $c_{1}$ and
$c_{2}$ are the distances of the center of mass of both links to
the corresponding joint axes. The length of the links are denoted
by $l_{1}$ correspondingly $l_{2}$. The moments of inertia of the
links, relative to their center of mass, are denoted by $I_{1}$
and $I_{2}$ respectively. Further, for the sake of simplicity, we
assume $l_{1} = l_{2} =l$, $c_{1} = c_{2} = 0.5l$, $I_{1} = I_{2}
= I$, and $m_{1} = 2m_{2} = m$. For the dynamics of rigid two-link
planar manipulator, described explicitly e.g. in
\cite{spong2006,Sicil2009}, and assumptions made above we obtain
the elements of $H$ matrix and $C$ and $G$ vectors as following:
\begin{eqnarray}
\nonumber   h_{11}        &=& ml^{2}\bigl(0.875 + 0.5 \cos q_{2}\bigr) +2I, \\
\nonumber   h_{12}=h_{21} &=& ml^{2}\bigl(0.25 + 0.5 \cos q_{2}\bigr) + I,\\
\nonumber   h_{22}        &=& 0.25 ml^{2} + I, \\
\nonumber   c_{1}         &=& -0.5 ml^{2} \sin q_{2}\bigl(2\dot{q}_{2}\dot{q}_{1} +\dot{q}_{2}^{2}\bigr),\\
\nonumber   c_{2}         &=& 0.5 ml^{2} \sin q_{2} \, \dot{q}_{1}^{2},\\
\nonumber   g_{1}         &=& m l g \bigl(1.5 \cos q_{1} + 0.5 \cos (q_{1} + q_{2})\bigr),  \\
  g_{2}         &=& 0.5 m l g \cos(q_{1} + q_{2}).
\end{eqnarray}
The parameters assumed for numerical simulation are listed in
Table \ref{tab:simparams}. Note that several nonlinear parameters,
like the shape factors of Stribeck curve and hysteresis, are
assumed to have the same values for both axes, this for the sake
of simplicity. Further we note that, for the sake of
comprehensibility, the parameters related to the joint
transmission are denoted with `deg' and not `rad' units as
otherwise.
\begin{table}[!h]
  \renewcommand{\arraystretch}{1.3}
  \caption{Plant parameters used in numerical simulation}
 \label{tab:simparams}
  \begin{center}
  \begin{tabular} {|p{2cm}|p{2cm}|p{2cm}|}
  \hline
  Parameter         & Unit            & Value   \\[0.05cm]
  \hline \hline
  $J$                 & kg m$^2$        & $[1,1]^T$   \\
  $I$                 & kg m$^2$        & 0.5 \\
  $g$                 & m s$^{-2}$      & 9.8 \\
  $m$                 & kg              & 10 \\
  $l$                 & m               & 0.5 \\
  \hline
  $F_c$               & Nm              & $[10,10]^T$ \\
  $F_s$               & Nm              & $[5,5]^T$ \\
  $B$                 & Nm s rad$^{-1}$ & $[1, 1]^T$ \\
  $V$                 & s rad$^{-1}$    & $[2, 2]^T$ \\
  $\mu$               & unitless        & $[-2, -2]^T$ \\
  $\gamma$            & unitless        & $[500, 500]^T$ \\
  \hline
  $D$                 & Nm s deg$^{-1}$ & $[1, 1]^T$ \\
  $K_1$               & Nm deg$^{-1}$ & $[300, 300]^T$ \\
  $K_3$               & Nm deg$^{-1}$ & $[50000, 50000]^T$ \\
  $w_i$               & unitless        & $[0.4, 0.4]^T$ \\
  $\psi$              & unitless        & $[300, 300]^T$ \\
  $\xi$               & unitless        & $[500, 500]^T$ \\
  $\eta$              & unitless        & $[1.5, 1.5]^T$ \\
  \hline
  \end{tabular}
  \end{center}
  \normalsize
\end{table}
The corresponding $f$-$\dot{\theta}$ friction and $\tau$-$\Delta$
hysteresis curves are visualized in Fig. \ref{fig:nonlincurves}
(a) and (b) correspondingly.

The following results are obtained within numerical simulation of
the robot plant according to (1), (2), (25) plus an additional
viscous joint damping $D$ so that $\tau = \chi(\Delta) + D
\dot{\Delta}$. Note that the latter is also required for
stabilizing the numerically implemented hysteresis (13)-(16) at
steeply changes of the torsion value, e.g. at higher steps of the
applied input torque. Furthermore, in order to render more
realistic conditions of the motor drive feedback control, the
manipulator plant signals $\theta_1$ and $\theta_2$ are outputted
through the 14 bit per revolution quantization blocks, that is
adequate for the common motor drive encoders.

\begin{figure}[!h]
\centering
\includegraphics[width=0.48\columnwidth, height=3.5cm]{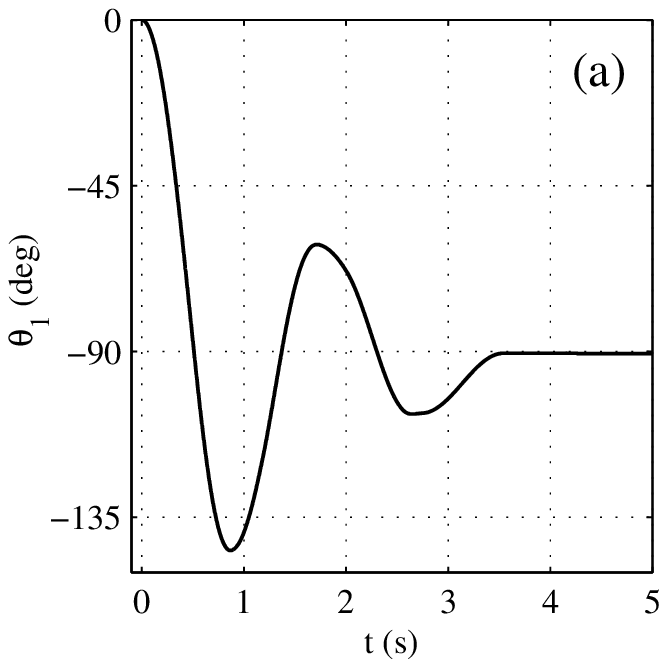}
\includegraphics[width=0.48\columnwidth, height=3.5cm]{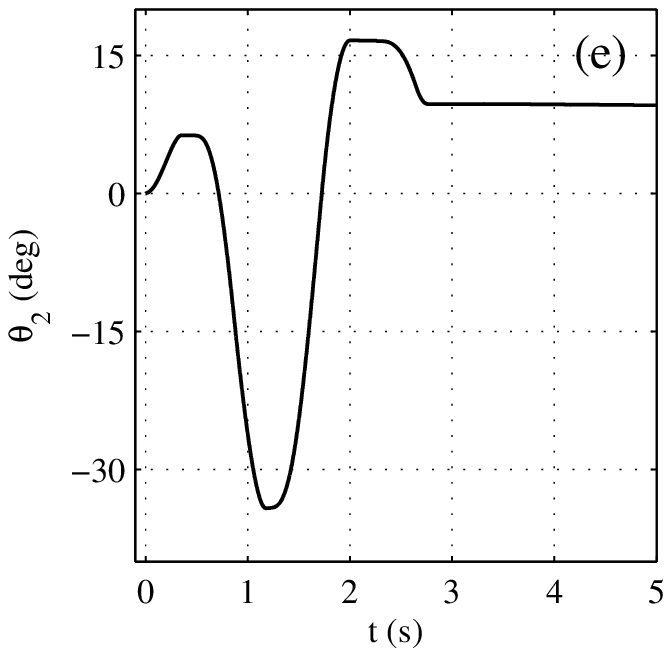}\\
\includegraphics[width=0.48\columnwidth, height=3.5cm]{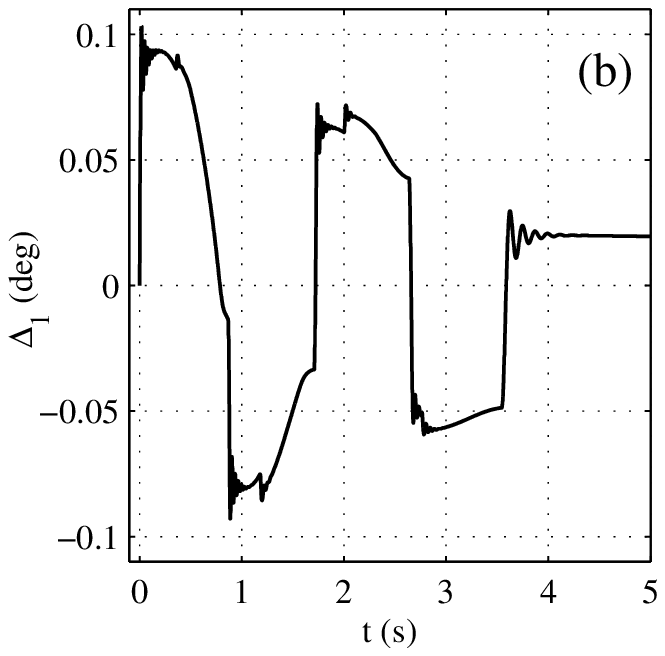}
\includegraphics[width=0.48\columnwidth, height=3.5cm]{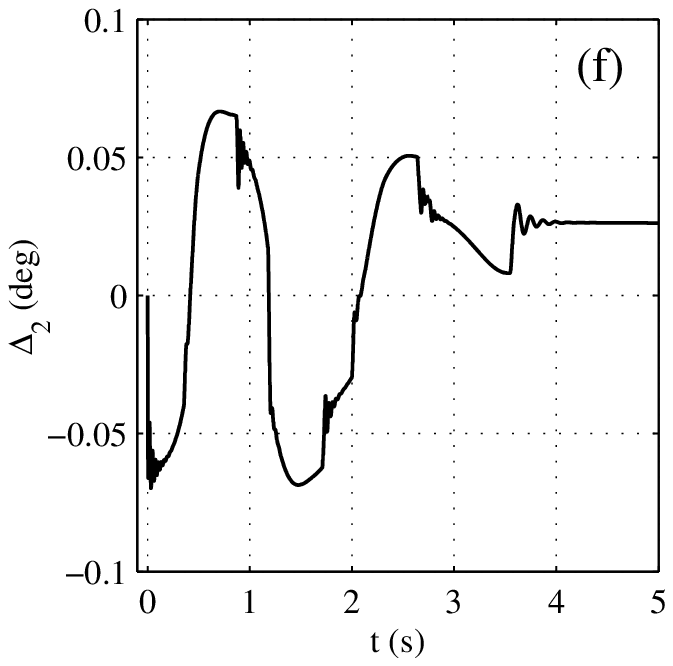}\\
\includegraphics[width=0.48\columnwidth, height=3.5cm]{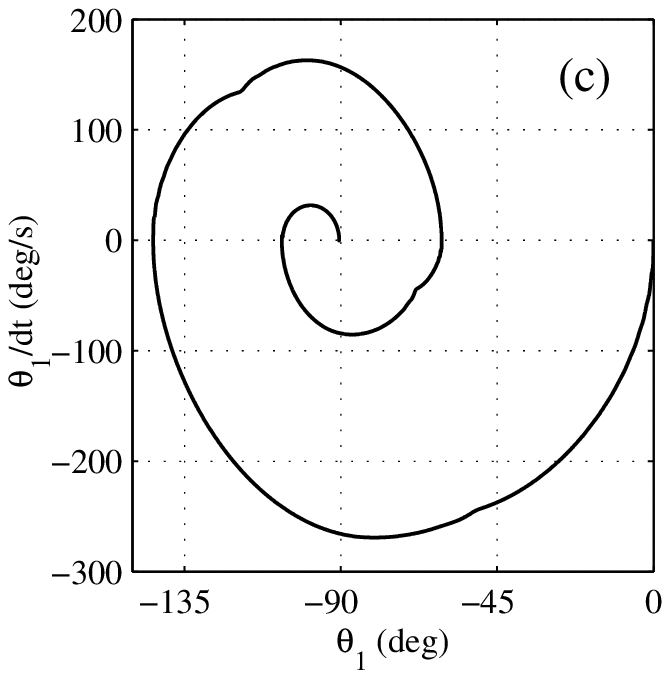}
\includegraphics[width=0.48\columnwidth, height=3.5cm]{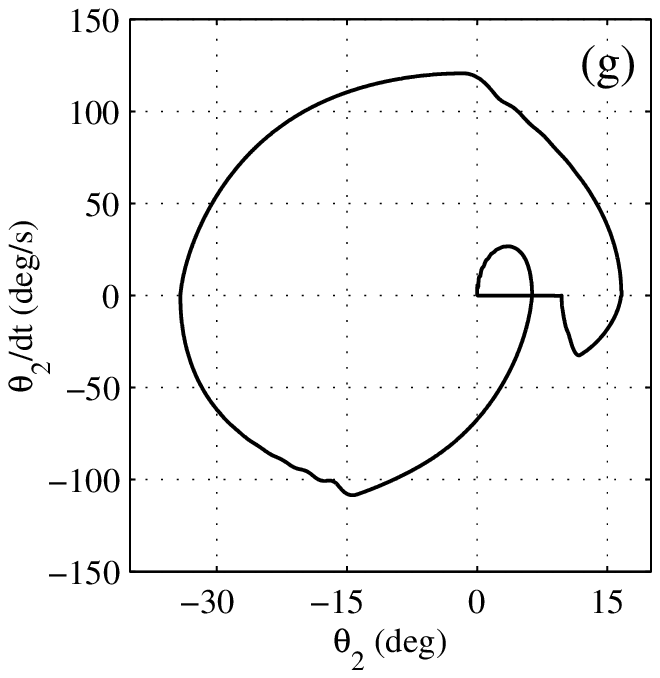}\\
\includegraphics[width=0.48\columnwidth, height=3.5cm]{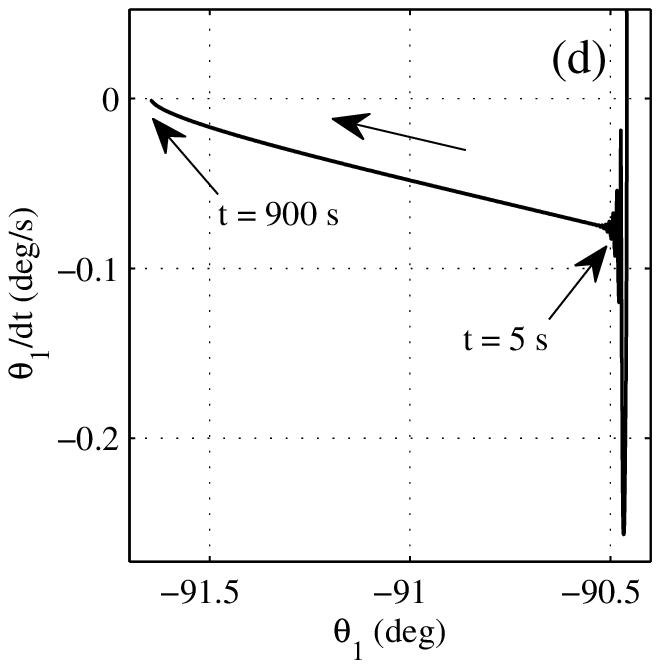}
\includegraphics[width=0.48\columnwidth, height=3.5cm]{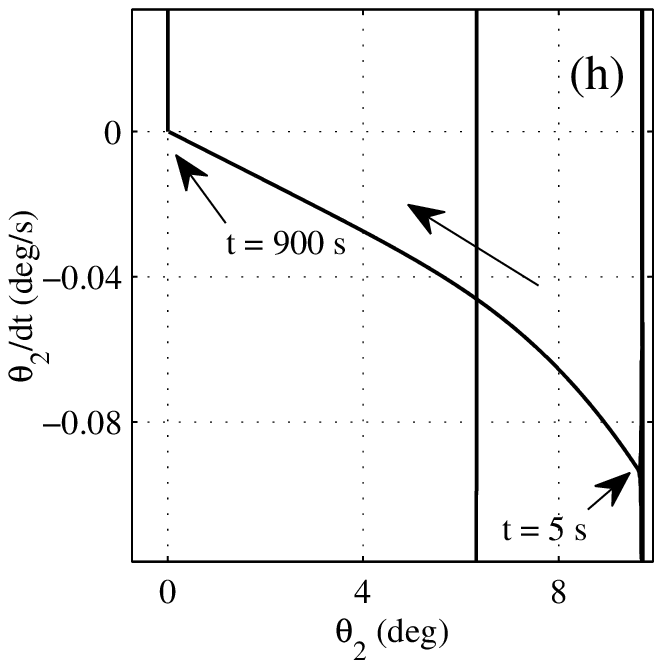}
\caption{Free fall response of two-link manipulator joints from
$\theta(t_{0}) = [0,\, 0]$ deg: 1st joint on the left and 2nd
joint on the right. (a), (e) motor drive position $\theta$; (b),
(f) joint torsion $\Delta$; (c), (g) state trajectories
$(\theta,\dot{\theta})$; (d), (h) zoom-in of
$(\theta,\dot{\theta})$ trajectories at slow relaxation dynamics.}
\label{fig:freefall}
\end{figure}
The effect of complex nonlinear dynamics of two-link planar
manipulator is visualized by means of a 'free fall' response. This
one assumes the initial joint configuration $\theta (t_{0})= [0,
0]$ deg which is the horizontal (outstretched) manipulator pose
with the maximal gravity acting on both axes. The free fall
response is shown in Fig. 5, for the 1st joint on the left and 2nd
joint on the right. The position trajectories are shown in Fig. 5
(a) and (e). Note that the second joint stops its large
displacement earlier than the first one, mainly due to the
friction. The micro-displacemens, visible in torsion response in
Fig. 5 (b) and (f) exhibit however nearly the same duration for
both joints. Remarkable is the fact of hysteresis lost motion
(non-zero torsion) at steady-state. The phase portrait in the
$(\theta , \dot{\theta})$ coordinates are shown in Fig. 5 (c) and
(g). The zoom-in of both trajectories in vicinity to final
equilibrium are depicted in Fig. 5 (d) and (h). One can see that
after the motion response on a 'fast' time scale, including both
the transients and 'quasi steady-states' up to about $t = 5$ s, an
extremely slow relaxation of nonlinear (creeping) dynamics occurs.
The final equilibrium is achieved after about $t = 900$ s while
the overall creeping displacement is about 1 deg for the 1st joint
and 9 deg for the 2nd one. The observed creeping phenomenon occurs
mainly due to an interplay between the hysteresis restoring torque
and friction torque nonlinearities. Here we note that no stiction
force is captured by (3) and (4) and, in reality, the joints may
stop without exhibiting a slow relaxation dynamics.

In the first approach (further denoted as `Control I'), the
full-order feed-forward control (8) is combined with the standard
PD feedback control so that the overall control law becomes
\begin{equation}\label{10}
u = K_{p}(\theta_r-\theta) + K_{d}(\dot{\theta}_r-\dot{\theta}) +
u_r\bigl(\theta_r\bigr),
\end{equation}
where the reference $\theta_r$ is obtained from $q_r$ through the
trajectory transformation as described in Section III.

In the second approach (further denoted as `Control II'), the
control law (17) is combined with the reduced-order
feed-forwarding os that
\begin{eqnarray}
\label{11}
  u &=& K_p e + K_d \dot{e} + K_p \tilde{\Delta}(\tilde{\tau}) + \tilde{u}_r, \quad \hbox{with} \\
\nonumber  \tilde{u}_r &=& \bigl(H(q_r) + J\bigr)\ddot{q}_r
+C(q_r, \dot{q}_r) +G(q_r) + f(\dot{q}_{r}).
\end{eqnarray}
Note that the feed-forward control part $\tilde{u}_r$ is not
explicitly accounting for joint elasticities. The setup control
gains used for both, Control I and Control II, plus the observer
gain are listed in Table \ref{tab:simcontrol}.
\begin{table}[!h]
  \renewcommand{\arraystretch}{1.5}
  \caption{Control parameters used in numerical simulation}
  \label{tab:simcontrol}
  \begin{center}
  \begin{tabular} {|p{2cm}|p{2cm}|p{2cm}|}
  \hline
  $K_p$ (Nm rad$^{-1}$)        & $K_d$  (Nm rad$^{-1}$s)               & $L$   \\[0.05cm]
  \hline \hline
  $[1.3,1.3]^T$     & $[0.43,0.43]^T$      & $[100,100]^T$\\
  \hline
  \end{tabular}
  \end{center}
  \normalsize
\end{table}

\begin{figure}[!h]
\centering
\includegraphics[width=0.49\columnwidth]{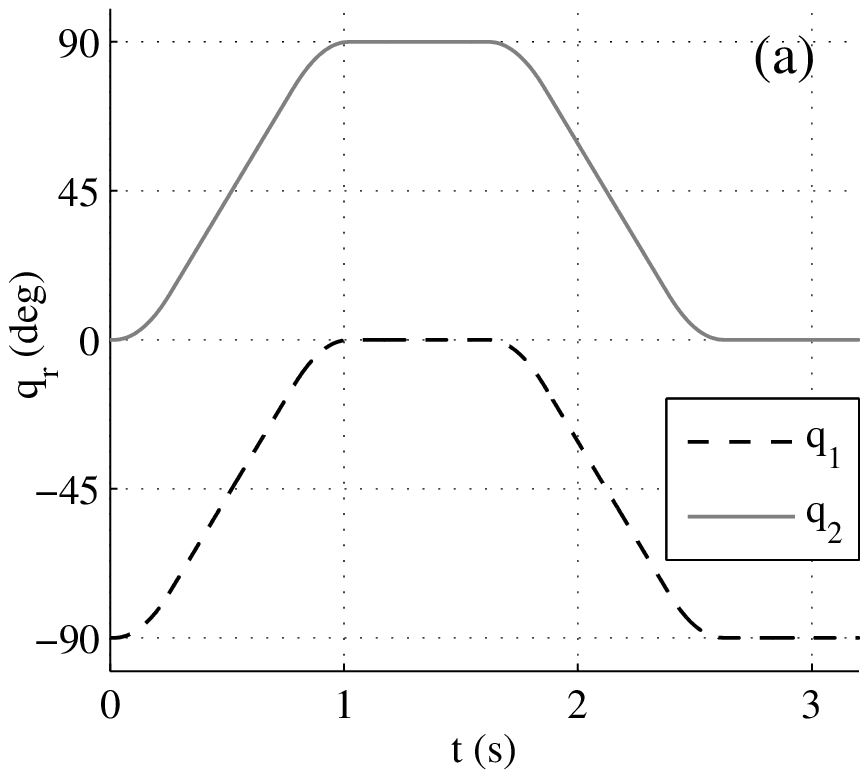}
\includegraphics[width=0.49\columnwidth]{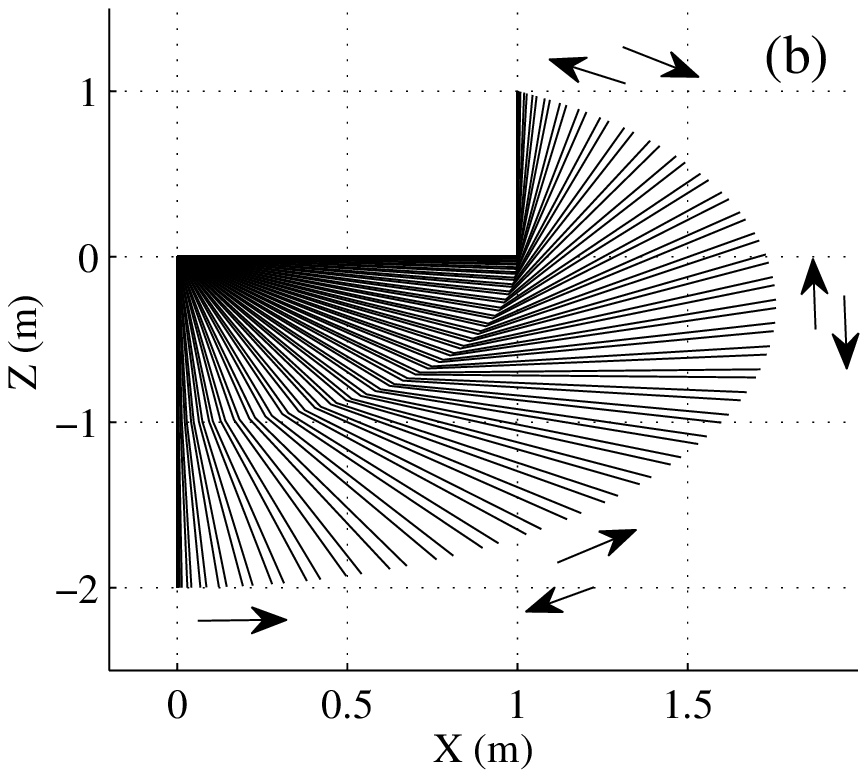}
\caption{Evaluated trajectory in the joint space (a) and Cartesian
space (b)} \label{fig:traject}
\end{figure}

The evaluated trajectory constitutes a simultaneous motion of both
links with the same shape of joint references $q_{(1,2),r} \in
\mathcal{C}^4$ as depicted in Fig. \ref{fig:traject} (a).
In order to visualize the motion of two-link manipulator in the
operational (Cartesian) space, and thus convey an impression about
the corresponding joint loads, we make use of the forward
kinematics of two-link planar manipulator according to
\cite{spong2006}. This is quite trivial and can be derived by
using a geometric approach as follows. Given the joint coordinates
$q_{1}$ and $q_{2}$ the Cartesian coordinates are given by
\begin{eqnarray}\label{eq:15a}
  X &=& l \cos q_{1} + l \cos(q_{1} + q_{2}), \\
  Z &=& l \sin q_{1} + l \sin(q_{1} + q_{2}).
  \label{eq:15b}
\end{eqnarray}
The stroboscopic motion of two-link planar manipulator with the
start configuration $q_{r}(t_0) = [-90, 0]$ deg and end
configuration $q_{r}(t_{3.2}) = [-90, 0]$ deg, coming through the
upper steady-state position $q_{r}(t_{1.1-1.6}) = [0, 90]$ deg, is
visualized in Fig. \ref{fig:traject} (b).

\subsection*{Control I} \label{sec:6:1}

The Control I has been evaluated on the reference trajectory shown
above by using three related configurations. First, the
feed-forward (8) only has been applied. Second, the feed-forward
has been augmented by the PD feedback as in (26), but without
reference position transformation, i.e. $\theta_{ r} = q_{r}$.
Third, the complete Control I as in (26) has been applied. The
control error of link positioning is shown in Fig. 7, for the 1st
joint in (a) and 2nd joint in (b). One can see that already the
single feed-forward control provides a relatively high positioning
accuracy, up to certain numerical (integrative) errors. This
argues in favor of the inverse dynamics computations (5)-(8) which
are implemented with a discrete-time and discrete-state solver and
are real-time compatible. The augmented feedback control, when
$\theta_{r} = q_{r}$, improves further the link position accuracy
but only up to the joint torsion values. The highest accuracy is
achieved in case of the full Control I as in (26). The accuracy is
close to hysteresis lost motion, see Fig. 4 (b).
Interesting fact is that the steady-state error at final
zero-gravity position (see Fig. 7 (a) for $t > 3$ sec) is inferior
for the full Control I comparing to the case when $\theta _{r} =
q_{r}$. This is because the reference value transformation does
not account for the actual hysteresis state which can be highly
varying during a feedback regulation in vicinity to the
torsion-torque origin. On the contrary, a dying-out oscillating
response of the Control I with $\theta _{r} = q_{r}$ apparently
drives the hysteresis to an erased (memory-free) state, thus
reducing the impact of hysteresis lost motion on the link
positioning error.
\begin{figure}[!h]
\centering
\includegraphics[width=0.98\columnwidth]{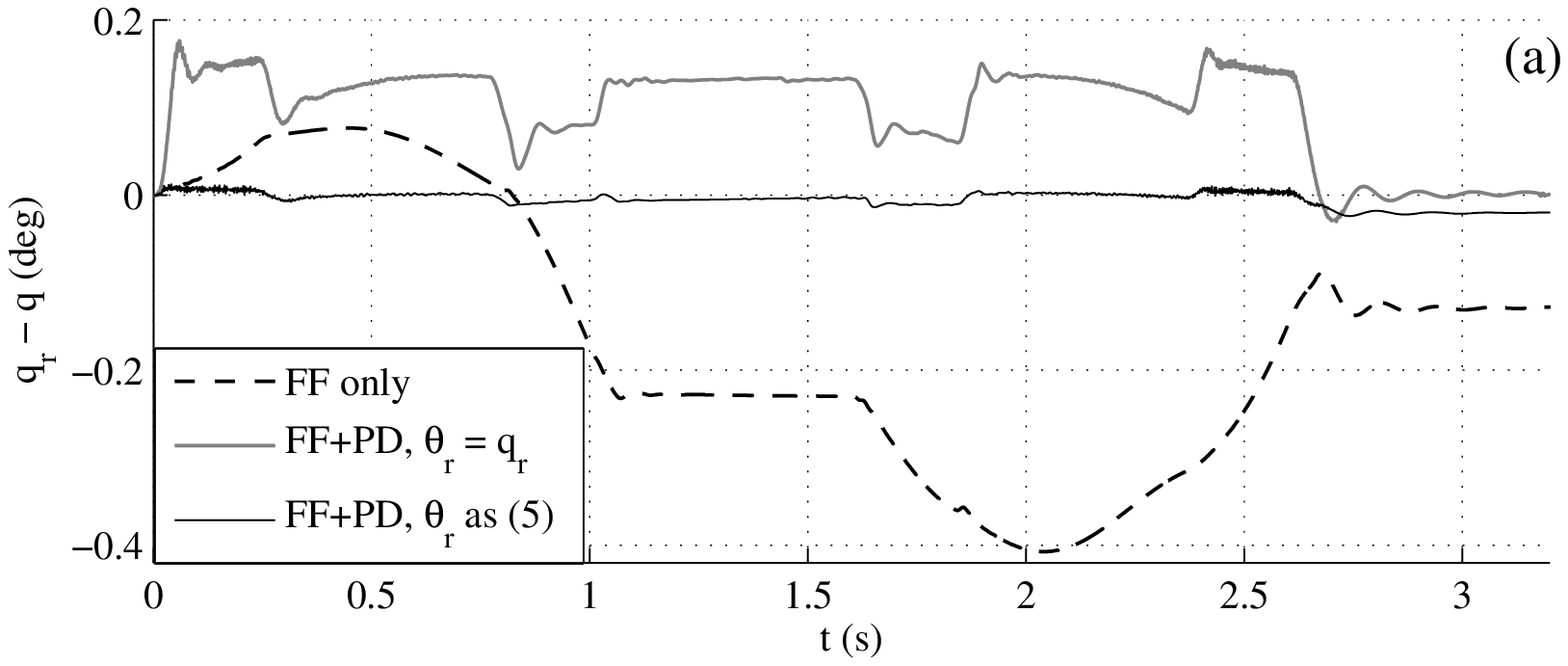}
\includegraphics[width=0.98\columnwidth]{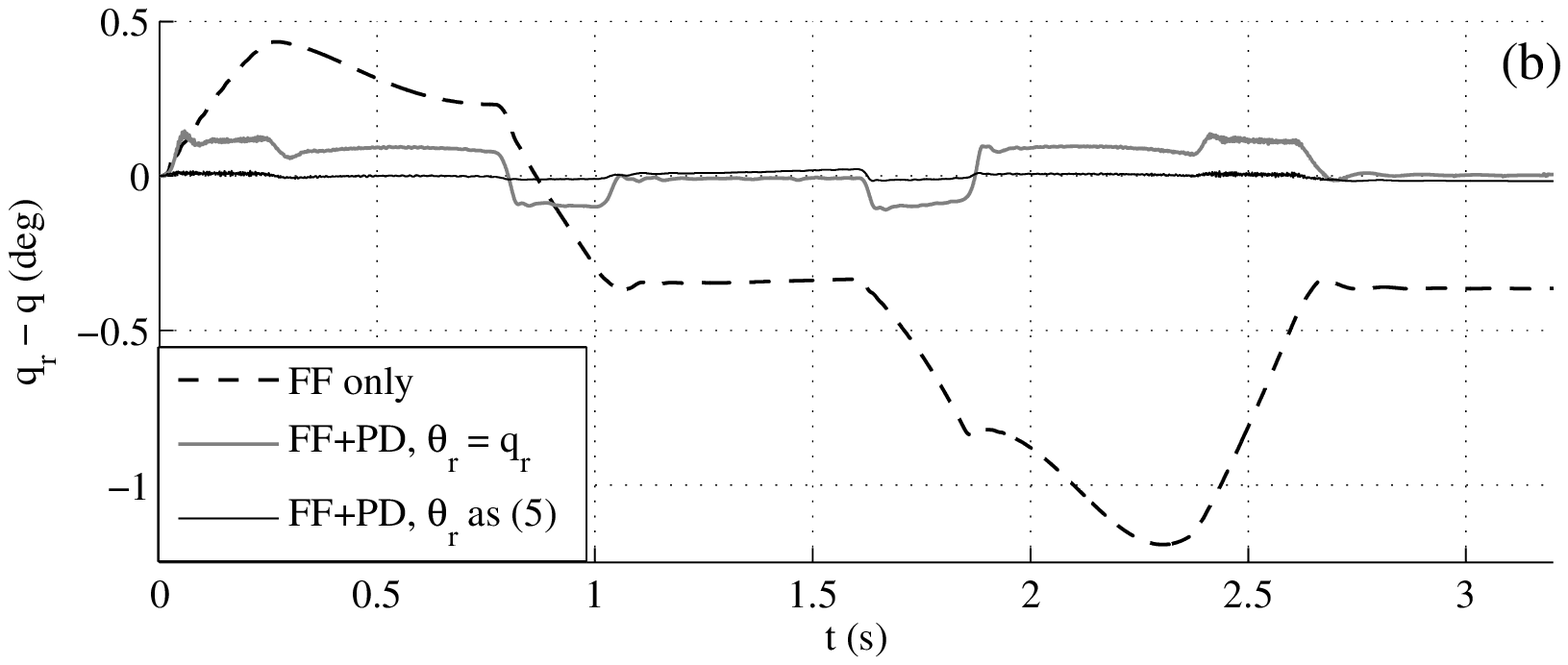}
\caption{Control error of the link position: 1st joint (a) and 2nd
joint (b)} \label{fig:control1error}
\end{figure}

\subsection*{Control II} \label{sec:6:2}

The Control II has been evaluated on the same reference trajectory
as before, also by using three related configurations for the sake
of comparison. First, only the feed-forward (FF) control
$\tilde{u}_r$ from (27) has been applied. Second, the feed-forward
control has been augmented by the same PD feedback control as
before (FF+PD), but with the reference $q_{r} = \theta $. Note
that this is equivalent to a standard rigid-manipulator control
(model-based feed-forward plus PD feedback) without considering
the joint elasticities. Third, the full Control II as in (27) has
been applied which incorporates the virtual sensor (FF+PD+VS).
Recall that the VS serves for predicting the actual joint torsion
as exemplary shown in Fig. 8 for the evaluated trajectory.
\begin{figure}[!h]
\centering
\includegraphics[width=0.48\columnwidth]{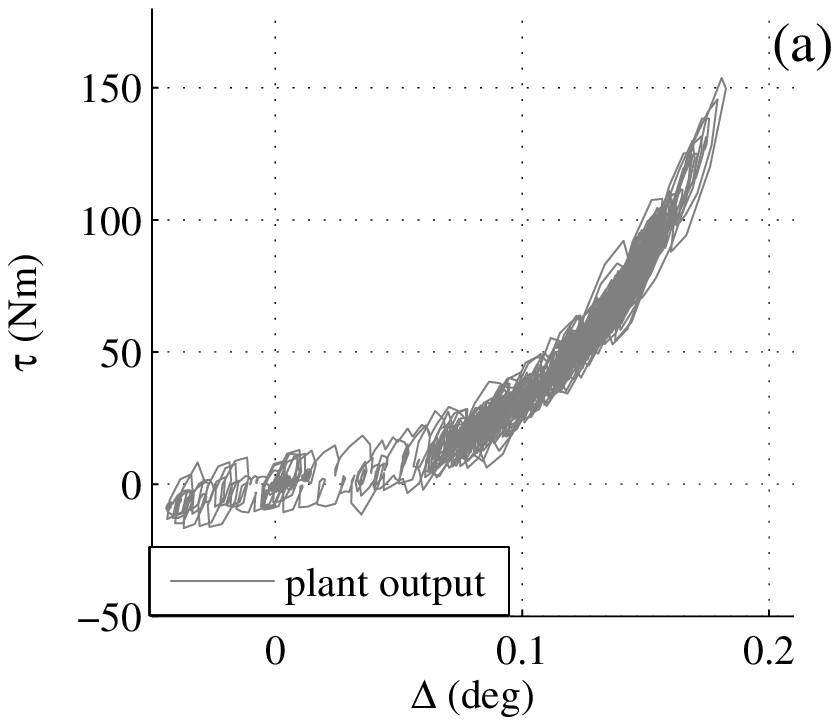}
\includegraphics[width=0.48\columnwidth]{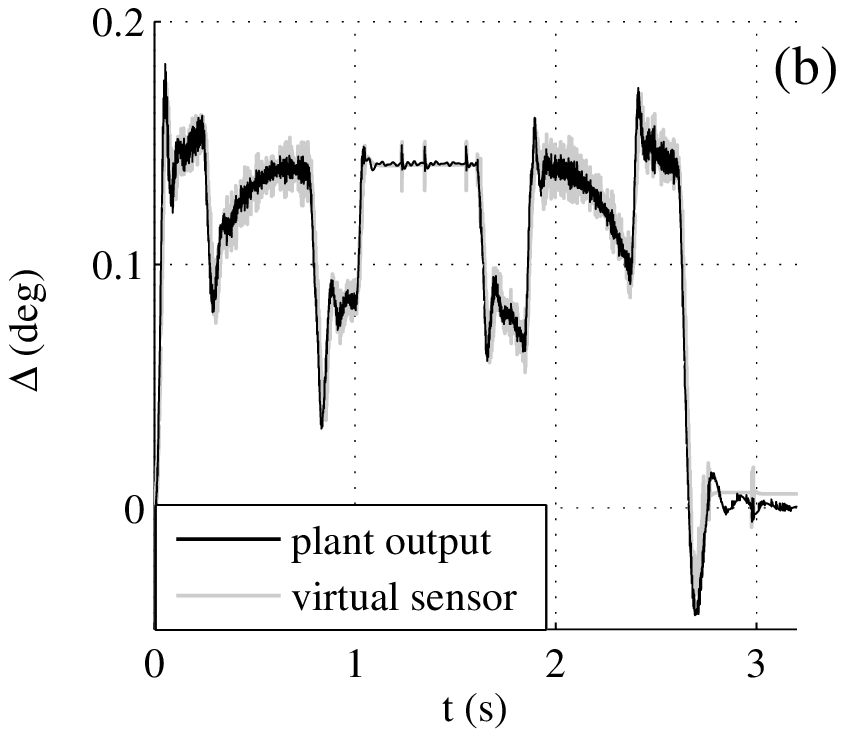}
\caption{Torsion behavior of the 1st joint during trajectory
control; torsion-torque portrait of the plant output (a),
comparison of the plant output with virtual sensor (VS) prediction
(b)} \label{fig:Control2_Hyst}
\end{figure}
The torsion-torque portrait, i.e. plant output, of the 1st joint
is shown in Fig. 8 (a). One can see a large set of minor
hysteresis loops inside of the major one. The relative torsion of
the 1st joint is compared with the VS prediction in Fig. 8 (b).
Both curves coincide well with each other for the transients and
steady-states and zero motion as well. The VS prediction offers a
slightly higher oscillating pattern which can be reduced by
decreasing the $L$ gains, however at costs of the slower
transients. The control error of link positioning is shown in Fig.
9, for the 1st joint in (a) and 2nd joint in (b). The FF+PD
control performance is comparable with that shown in Fig. 7 for
the Control I. The single FF control is, however, inferior
comparing with Fig. 7, since FF in (27) constitutes a reduced
feed-forwarding, i.e. without accounting for joint elasticities.
The best link positioning accuracy is achieved with the FF+PD+VS
control which compensates for the actual joint torsion. At the
same time, the FF+PD+VS is inferior to the full Control I, compare
Figs. 7 and 9. This is quite natural since the predicted torsion
value, even when accurate enough, solely enters the proportional
feedback control term and thus can be just as efficient as the
corresponding proportional control part is. An increase of the
$K_{p}$ gain can further improve the steady-state performance of
FF+PD+VS control, however, at costs of the higher transient
overshoots. At the same time, one can notice that the Control II
can better cope with hysteresis lost motion at zero gravity
steady-state, compare Figs. 7 (a) and 9 (a) at time $t > 3$ s.
This is quite natural since the concept of VS deals with
estimating the actual hysteresis joint torsion independent of the
reference trajectories and modeling of manipulator dynamics.
\begin{figure}[!h]
\centering
\includegraphics[width=0.98\columnwidth]{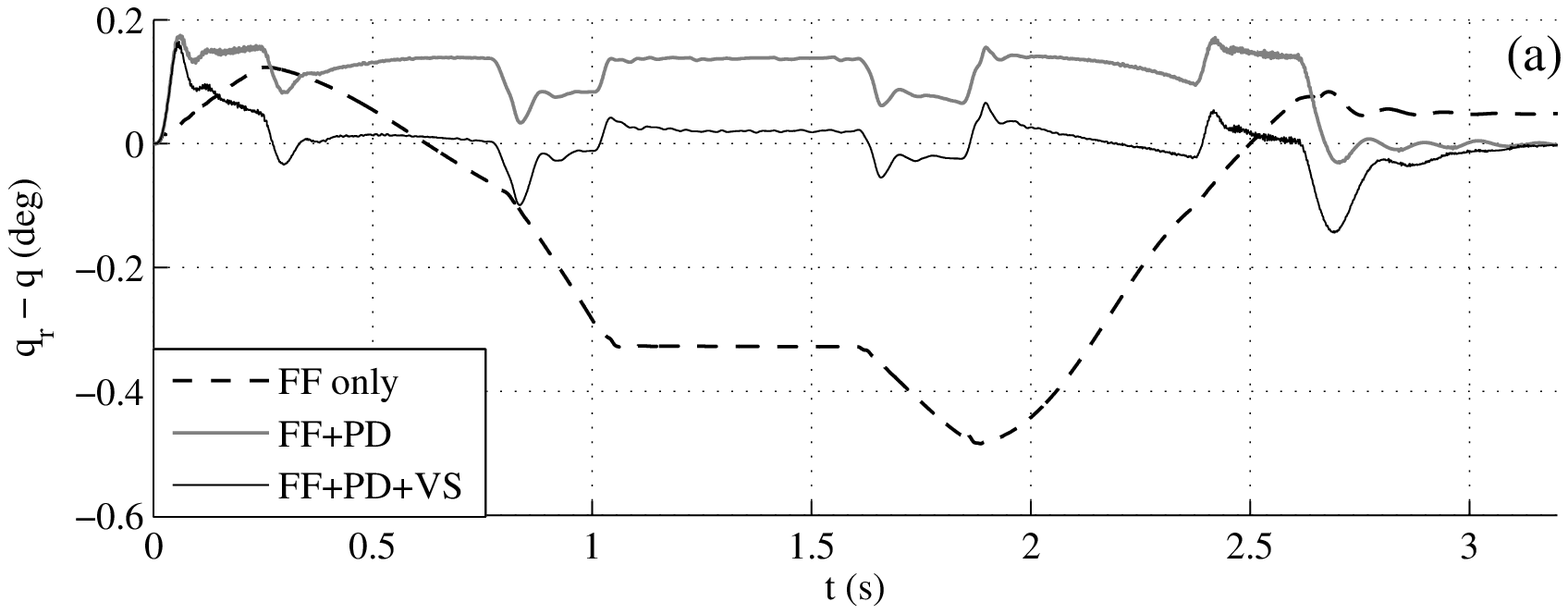}
\includegraphics[width=0.98\columnwidth]{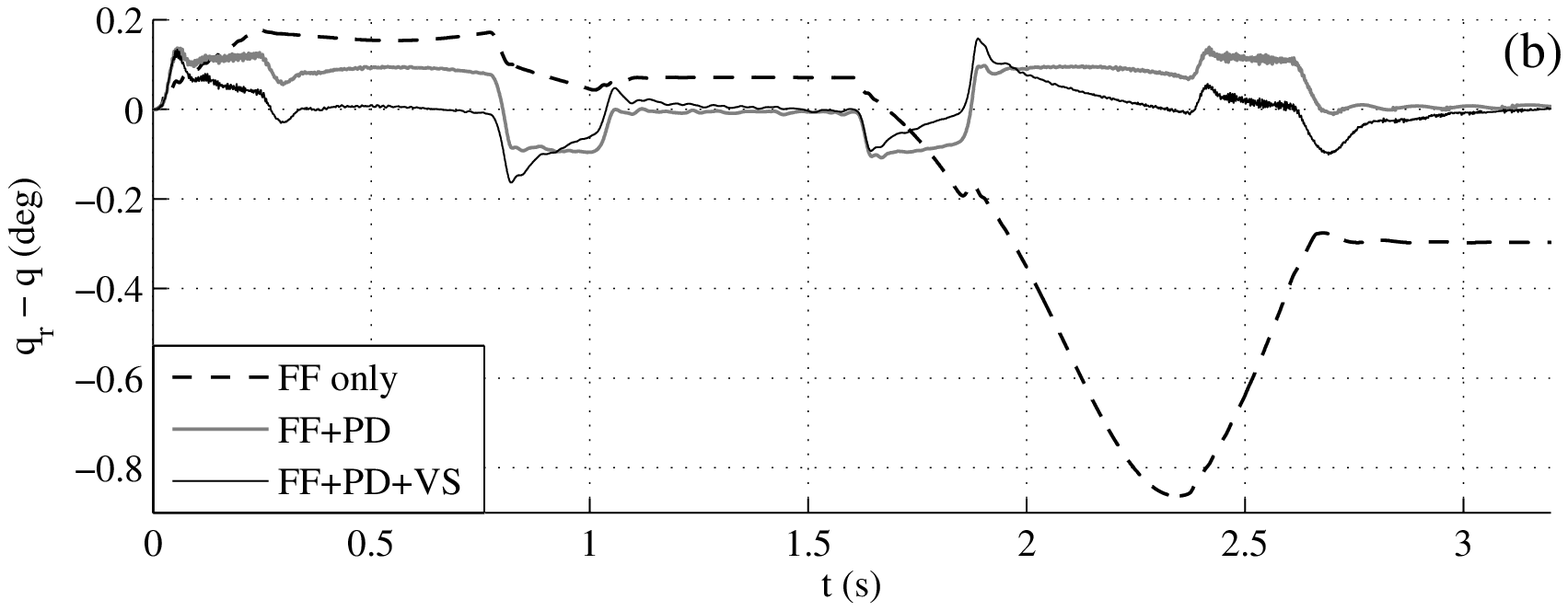}
\caption{Control error of the link position: 1st joint (a) and 2nd
joint (b)} \label{fig:control2error}
\end{figure}

\subsection*{Discussion} \label{sec:6:3}

Based on the description of both control methods given in Sections
II and IV and control evaluation made above the following remarks
can be drawn.

(i) Both methods, Control I and Control II, are the approaches to
compensate nonlinear torsion that should be useful when high
positioning accuracy is required. This compensation, however, in
order to be effective requires an accurate modeling of the motor
and gear, including friction effects. Both control methods are
suitable for a multi-link robotic manipulator and account for the
related multiple-input-multiple-output plant with cross-couplings.
Both control methods require the nonlinear (hystersis)
torsion-torque map of each flexible joint to be identifiable and
thus available as an accurate model.

(ii) The Control I provides in total the best accuracy of link
positioning. At the same time, the control accuracy becomes
inferior at zero gravity steady-state, where the hysteresis
trajectories, close to the torsion-torque origin, give rise to
residual link positioning errors. These cannot be compensated by
the Control I since the latter relies on a feed-forward
computation of torsion trajectories and thus does not account for
actual state of the joint torsion. Another flaw of Control I is
that this requires an accurate model of the multi-link robotic
manipulator, including the configuration-dependent inertia and
gravity terms and friction as well, see equation (8). These can
change, however, during the manipulator's operation, e.g. due to
an additional pay-load applied on the end-effector.

(iii) The Control II is slightly interior at compensating the
relative joint torsion, comparing to the Control I. Its
performance is directly related to that of the underlying PD
feedback control and to accuracy of predicting the actual joint
torsion. The accuracy of torsion's prediction constitutes a
trade-off between the fast transients of joint torque estimate and
high-frequent (chattering) components at steady-states and
settling from the transients. At the same time, the Control II
does not require an accurate model of the multi-link robotic
manipulator and accounts for the actual joint torsion state at
hand. The utilized concept of virtual sensor of the joint torsion
requires, however, an accurate modeling of motor drive friction
and motor drive inertia. While the motor drive inertia can be
assumed as constant, i.e. time-invariant, the friction can
underlie large uncertainties due to e.g. thermal effects, wear,
dwell time, and others. The uncertain friction behavior and its
observation and compensation have been recently addressed in
\cite{Ruder2015f}.

\section{CONCLUSIONS}
\label{sec:6}

In this paper, we have analyzed and compared two control
approaches aimed at compensating for the nonlinear torsion in
flexible joint robots. Both approaches differ in requirements
posed on the available models of system dynamics and their
accuracy. The first method assumes an accurate model of a
rigid-link manipulator dynamics and its inverse. Based thereupon
the reference trajectory is transformed from the link space into
that of the motor drives, which are under a closed-loop control.
The PD feedback control is combined with the full-order reference
torque feed-forwarding. The second method relies on an accurate
model of the motor drives and allows for observing the reactive
joint torque based on the generalized momenta. The observed
reactive joint torque allows for computing the relative joint
torsion. Thus, the motor drive feedback control operates in the
'virtual' joint link space by accounting for torsion. Both
approaches make use of the same torsion-torque hysteresis map and
its inverse. It turns out that depending on the model availability
and control specification each of the methods can offer several
assets and drawbacks. The simulation example of a two-link planar
manipulator under gravity showed the applicability of both control
methods.


\bibliographystyle{IEEEtran}
\bibliography{references}

\end{document}